\documentclass[pra,preprint,showpacs]{revtex4}
\usepackage[centertags]{amsmath}
\usepackage{amsfonts}
\usepackage{amssymb}
\usepackage{amsthm} 
\usepackage{newlfont}
\usepackage{epsfig}
\usepackage{amscd}
\usepackage{graphicx}
\usepackage{epsfig}
\usepackage{footnote}
\usepackage{lipsum}
\usepackage{color}
\usepackage{xcolor}
\usepackage{graphicx}
\usepackage{multirow}
\usepackage{subfig}
\usepackage{amscd}

\newcommand{\beq}{\begin{equation}}
\newcommand{\eeq}{\end{equation}}
\newcommand{\ba}{\begin{array}}
\newcommand{\ea}{\end{array}}
\newcommand{\bea}{\begin{eqnarray}}
\newcommand{\eea}{\end{eqnarray}}
\begin{document}

\begin{center}
{\large \sc \bf {Operations with elements 
of transferred density matrix via unitary transformations on extended receiver }
}

\vskip 15pt

{\large 
A.I.Zenchuk 
}

\vskip 8pt

{\it 
Institute of Problems of Chemical Physics RAS,
Chernogolovka, Moscow reg., 142432, Russia}

\end{center}

\begin{abstract}
We combine the long-distance quantum state transfer and simple operations with the elements of the transferred (nor perfectly) density matrix. These operations are 
turning some matrix elements to zero,  rearranging the matrix  elements and preparing their linear combinations with required coefficients.  The basic tool performing these operations is the unitary transformation on the extended receiver.  A system of linear algebraic equations can be solved in this way as well.  Such operations are numerically simulated on the basis of 
42-node spin-1/2 chain with the two-qubit sender and receiver.
\end{abstract}

\maketitle

\section{Introduction}
\label{Section:Introduction}

The development of quantum counterparts of classical algorithms is a beneficial direction in quantum information processing. 
Since unitary transformations play a fundamental role in quantum mechanics, the problem of their implementation and control is of general interest. A principal role plays a simplest two-qubit operator called controlled-NOT (CNOT) which entangles states of two qubits. This operation was generalized to controlled-$U$ operator   ($C^n(U)$) with an arbitrary one-qubit unitary transformation $U$ controlled by $n$ ($n\ge 1$) qubits  \cite{BBCVMSSSW,NCh,LLS}. Selecting a proper combination of unitary operators (perhaps, combined with certain measurements) allows one to realize various algorithms useful in a wide range of applications.
To such algorithms we refer the quantum Fourier transform (QFT) \cite{NCh}, Hamiltonian simulation \cite{BACS,Ch} and phase estimation \cite{NCh,LP,CEMM}. These algorithms  are used as subroutines in other
algorithms such as the HHL-algorithm for solving  linear systems \cite{HHL,WZRL},
 the algorithms for solving systems of nonlinear equations \cite{QH} and for 
quntum realization of elementary matrix operations using Trotter product formula \cite{ZZRF}.
The HHL-algorithm was   first performed in photon systems \cite{CWSCGZLLLP,BKRLDAW} and  recently  it was  implemented on the basis of superconducting qubit system to solve 
 the system of two linear equations  \cite{ZSCXLG}. The HHL-algorithm is  also effectively used in the quantum algorithm for data fitting over the exponentially large data set \cite{WBL}. 
Review on quantum machine learning is represented in \cite{BWPRWL}. The problem of quantum speedup, which is crucial for justifying the advantages of quantum computations,   is discussed in the above references as well. Basic quantum algorithms and their implementations in quantum information theory can be found in book \cite{NCh}.  
The research on realization of quantum gates goes further to work out the effective remote control of unitary operations (operator teleportation) \cite{HPXLG,QZACB} with  multi-qubit  controlled-phase gates \cite{WY} among them. 

In our paper we consider the remote preparation of initial state at one side of the communication line (sender) 
which after been (non-perfectly) transferred to the (extended) receiver is subjected to the unitary transformation with the purpose of obtaining the desired state. 
In other words, we combine the long-distance non-perfect mixed state transfer and the operations with the elements of  the transferred density  matrix. Thus, the communication line includes the sender (where the initial state is to be prepared), the receiver (where the final state is to be detected) the extended receiver (which includes receiver and several additional nodes (ancilla) and serves as a platform for applying the unitary transformation aimed to perform  required operations) and the  transmission line (connecting the sender with the extended receiver).   
 During the evolution due to the  spin-spin interactions, each element of the receiver's density matrix becomes, in general,  a linear combination of all the elements of the initial sender's density matrix. This phenomenon is an obstacle for the  state transfer 
but not for the 
remote manipulations with  the matrix elements as is shown further in this paper. To (partially) remove this mixture 
 of elements  we chose the evolution   preserving the number of excitations in the system \cite{FZ_2017}. This constraint   provides the independent evolution of the multiple-quantum (MQ) coherence matrices, so that the matrix elements mix only inside of each  MQ-coherence matrix of a particular order. 
 {We recall that the $n$-order  MQ-coherence matrix collects those elements of the density matrix which correspond to the state-transitions changing the $z$-projection of the total spin by $n$.}
 The importance of such  independent evolution of MQ-coherence matrices is associated with different meaning of the matrix elements assembled inside of each coherence matrix. For instance, the diagonal elements (a part of the zero-order coherence matrix) represent the probabilities of state realization and therefore  must be nonnegative. In addition, their sum equals one. 

We deal with the  spin chain  considered in Ref.\cite{Z_2018} { where  the unitary transformation of the extended receiver was used for the
structural restoring of the nondiagonal part of the transferred matrix. We show that such transformation can be an effective tool for }
performing simple   operations  with a mixed state of a quantum system.  
{Of course, the problem of practical realization of required unitary transformations is directly related to such approach. In this regards we, first of all,  relay on the so-called Solovay-Kitaev theorem \cite{NCh}. According to this theorem, any unitary transformation can be constructed using one-qubit rotations and CNOTs (simplest two-qubit operations). In addition, the operator formation is a central point of the duality computing  \cite{Long,LY,LYCh,G}. In this method, the quantum system consists of two parts. The first one is  the internal subsystem which serves for computing. The second part is the (multi-qubit) ancila, on which two new quantum gates are realized: quantum wave divider and quantum wave combiner. Using  controlled unitary operations and controlled measurement acting on the internal subsystem  one can construct  various non-unitary (in general) operators acting on the internal subsystem (allowable generalized quantum gate). Realization  of  such gates is discussed in \cite{ZCL}.
}

{ As  operations with the elements of the transferred density matrix we,} consider the following  manipulations.
\begin{itemize}
\item
{\it Structural restoring of the nondiagonal part of the density matrix  supplemented with 
turning some nondiagonal matrix elements to zero.}
In this case, each  nondiagonal element multiplied by its scale factor maps into the  appropriate nondiagonal element of the receiver's density matrix thus establishing the structural restoring \cite{Z_2018}. However, there is a group of  elements having zero scale factors. 
\item
{\it Rearranging matrix elements.} Unlike the usual structural restoring, some  nondiagonal elements of the sender's density matrix 
multiplied by the scale factor exchange their positions at the receiver's state. 
\item
{\it Constructing certain linear combinations of matrix elements.} This operation results in
the linear combination (with needed coefficients) of several nondiagonal
elements of the sender's density matrix and assigns this combination to a proper 
element of the receiver's density matrix.   
\item
{\it Solving systems of linear algebraic equations.} Unlike the HHL-algorithm, our protocol is based on the mixed states.  Certain elements of the sender's density matrix are used as the input data (the right-hand side of the linear system). Then the evolution following by the unitary transformation on the extended receiver  makes such linear combinations of these elements that 
equals to the  solutions of the considered  linear system and place the result in the appropriate elements of the 
receiver's density matrix. 
\end{itemize}

In all these manipulations, we use the sender's state as the input information, the receiver's state as the 
output information and the unitary transformation 
of the extended receiver as a tool allowing to manipulate the  elements of the output state.
The transmission line is an optional part, it connects the sender with the extended receiver and serves to perform the remote manipulation with the input data. It can be absent if we are interested in the local manipulations. We emphasise that our protocols  do not include a classical communication channel and therefore can be referred to completely quantum processes.

The paper is organized as follows. 
In Sec.\ref{Section:instev}, we discuss the general protocol of operations  with the transferred density matrix, including structure of the communication line, the initial state and the general form of the evolution operator. The set of  manipulations with the receiver's density matrix elements is introduced in Sec.\ref{Section:manipulations}. In Sec.\ref{Section:model},
we briefly describe a particular model   of communication line used in Sec.\ref{Section:numerics} 
for numerical simulations of operations with density matrix elements. The paper concludes with Sec.\ref{Section:conclusions}. The general structure of the unitary transformation and examples of such transformation realizing the operations with matrix element 
are represented in the Appendix, Sec.\ref{Section:appendix}.

\section{Initial state and evolution operator}
\label{Section:instev}
\begin{figure*}
\epsfig{file=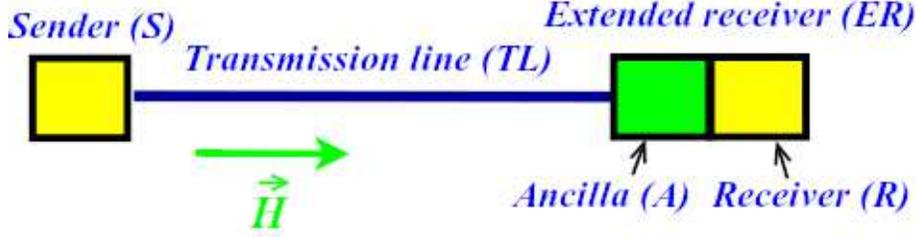,
  scale=3
   ,angle=0
}  
\caption{Spin-1/2 communication line consisting of the sender (S), transmission line (TL), ancilla (A) and  receiver (R). The extended receiver combines     the  ancilla and receiver. 
}
  \label{Fig:CL} 
\end{figure*}

\begin{figure*}
\epsfig{file=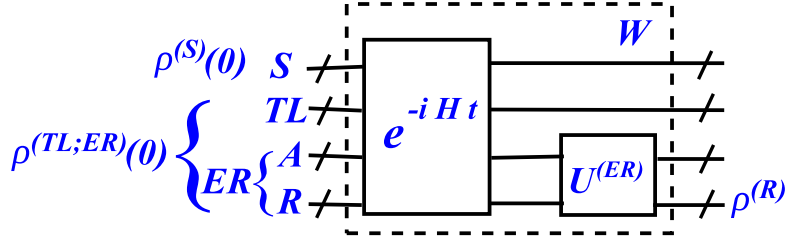,
  scale=0.5
   ,angle=0
}  
\caption{General scheme  manipulating the matrix elements of $\rho^{(R)}$.  The initial state is (\ref{tpIS}), the evolution operator $W$ 
consists of two part: the evolution of the whole system under the Hamiltonian $H$ followed by   the local unitary transformation $U^{(ER)}$ of the extended receiver. The finial state $\rho^{(R)}$ is detected at the receiver.}
  \label{Fig:EV} 
\end{figure*}

We consider the spin-1/2 communication line \cite{Z_2018} which includes the  sender ($S$), transmission line ($TL$), receiver ($R$) and ancilla ($A$)   in the strong external  magnetic field directed along the chain ($z$-direction), Fig.\ref{Fig:CL}.   The receiver and ancilla form the extended receiver ($ER$).
The general protocol  performing  certain manipulations with  nondiagonal elements of $\rho^{(R)}$ is schematically represented in Fig.\ref{Fig:EV}.
 First of all, we start with the tensor-product initial state
\begin{eqnarray}\label{tpIS}
\rho(0)=\rho^{(S)}(0)\otimes\rho^{(TL,ER)}(0),
\end{eqnarray}
where $\rho^{(S)}$ is the density matrix of the sender $S$, and $\rho^{(TL,ER)}$ is the density matrix of the subsystem $TL\cup ER$.
Let both the sender and receiver have the same dimensionality. 
If $W(t)$ is the evolution operator describing  dynamics of the given quantum system, then the  state of the receiver at some time instant $t$ reads
\begin{eqnarray}\label{gen}
&&
\rho^{(R)}_{N^{3};M^{3}}(t) = 
\sum_{I^{1},J^{1}} T_{N^{3}M^{3};I^{1}J^{1}}(t)\rho^{(S)}_{I^{1}J^{1}}(0),\\\label{genT}
&&
T_{N^{3}M^{3};I^{1}J^{1}}=\sum_{N^{1}N^{2}} \sum_{I^{2},I^{3},J^{2},J^{3}} W_{\{N\};\{I\}}  
\rho^{(TL,ER)}_{I^{2}I^{3};J^{2}J^{3}} W^+_{\{J\};N^{1}N^{2}M^{3}},
\end{eqnarray}
where we use the short notation  $\{N\}$ for $(N^1N^2N^3)$ and similar notation   $\{I\}$, $\{J\}$.
The operator $T$ in this formula maps the sender's initial state into the receiver's state.
Here the capital latin multi-indexes with superscripts 1, 2 and 3 are associated, respectively, with the sender, the transmission line joined with the ancilla, and the  receiver. These indexes include set of elements each associated with the particular spin of the appropriate subsystem and can take either 0 or 1. Thus, for the $N$-qubit communication line with the two-qubit sender and receiver we write
\begin{eqnarray}
I^{k}=\{i^k_1 i^k_2\},\;\;k=1,3,\;\;I^{2}=\{i^2_1\dots i^2_{N-4}\}.
\end{eqnarray}
Now we discuss the structure of the evolution operator $W$ including two blocks as shown in Fig.\ref{Fig:EV}.
First of all, it includes the  spin dynamics described by some Hamiltonian $H$: 
$V(t)=e^{-i H t}$.
In addition, the evolution operator $W$ includes the unitary transformation $U^{(ER)}$ of the extended receiver. This transformation
involves the set of additional parameters $\varphi_i$ (hereafter we call them $\varphi$-parameters) needed to control the state of the receiver. Then
\begin{eqnarray}\label{W}
W=(E^{(S,TL)}\otimes U^{(ER)})  V(t) .
\end{eqnarray}
Here $E^{(S,TL)}$  is the identity operator in the state space of the subsystem $S\cup TL$.

Hereafter we deal with such Hamiltonian $H$ and unitary transformation $U^{(ER)}$  that conserve the excitation number during the spin evolution, i.e.,
\begin{eqnarray}\label{comm}
[H,I_z]=0,\;\;[U^{(ER)},I^{(ER)}_z]=0,
\end{eqnarray}
where $I_z$ and $I_z^{(ER)}$ are  the $z$-projections  of the total spin momentum of, respectively, the whole communication line and the subsystem $ER$. 
Then, according to Ref. \cite{FZ_2017}, the MQ-coherence matrices of the whole system  do not interact during such evolution. 
In addition, to transfer the MQ-coherence matrices from the sender $S$ to the receiver $R$ 
without mixing, the initial density matrix $\rho^{(TL,ER)}(0)$ must include  only  the zero-order coherence matrix \cite{FZ_2017}. In particular,  it can be  the  diagonal matrix  $\rho^{(TL,ER)}(0)$ (for instance, the thermodynamic equilibrium state). Then eq.(\ref{genT}) becomes
\begin{eqnarray}\label{T}
T_{N^{3}M^{3};I^{1}J^{1}} = \sum_{N^{1}N^{2}} \sum_{I^{1},I^{2},I^{3},J^{1}} W_{N^{1}N^{2}N^{3};I^{1}I^{2}I^{3}} \rho^{(TL,ER)}_{I^{2}I^{3};I^{2}I^{3}}(0) W^+_{J^{1}I^{2}I^{3};N^{1}N^{2}M^{3}}.
\end{eqnarray}
Commutativity (\ref{comm}) implies  the following constraint for the evolution operator $W$:
\begin{eqnarray}
 W_{N^{1}N^{2}N^{3};I^{1}I^{2}I^{3}}=0 \;\;\;{\mbox{if}}\;\;\;\sum_{i=1}^3|N^{i}| \neq \sum_{i=1}^3|I^{i}|,
\end{eqnarray}
where $|N^{i}|$ and $|I^{i}|$ mean the sum of entries of the multi-indexes $N^{i}$ and $I^{i}$. We also set zero the energy of the ground state (the state without excitations), which means  
\begin{eqnarray}
W_{0^10^20^3;0^10^20^3} =1,
\end{eqnarray}
where $0^i$, $i=1,2,3$,  are the multi-indexes with all zero entries.

\section{Manipulations with receiver's density matrix elements}
\label{Section:manipulations}
{Before proceed to the numerical simulation of the particular operations with matrix elements in Sec.\ref{Section:numerics}, we describe general structures of these operations in terms of elements (\ref{T}) of the $T$-operator. This operator includes the $\varphi$-parameters of the unitary transformation $U^{(ER)}$, which
allow to  control the structure of the receiver's density matrix elements. 
In particular, the following operations with elements inside of $n$-order coherence matrix 
are of our interest, where we assume that the time instant is properly fixed (for instance, as in Sec.\ref{Section:model}).}

1. {\it Structural restoring of initial  sender's  state \cite{Z_2018}.} In this case, by choosing the $\varphi$-parameters, we have to satisfy the following constraints:
\begin{eqnarray}\label{restore}
&&
T_{N^3M^3;I^1J^1} = \lambda_{N^3M^3} \delta_{N^3 I^1}\delta_{M^3 J^1},\;\;(N^3, M^3)\neq (0^3, 0^3),\\\label{restore_norm}
&&
T_{0^3  0^3; I^1J^1}=  \delta_{I^1J^1} (1-\lambda_{I^1I^1}), \;\; (I^1, J^1)\neq (0^1, 0^1),\\\label{restore_norm2}
&& T_{0^3  0^3; 0^1 0^1}= 1.
\end{eqnarray} 
As the result, the structurally restored state reads  
\begin{eqnarray}\label{rho_restore}
&&
\rho^{(R;n)}_{N^3;M^3}=\lambda_{N^3M^3}\rho^{(S;n)}_{N^3;M^3},\;\;(N^3, M^3)\neq (0^3, 0^3),\\\label{rho_restore_norm}
&&
\rho^{(R;n)}_{ 0^3; 0^3} =1- \sum_{I^1 \neq 0^1} \lambda_{I^1I^1}  \rho^{(S;n)}_{I^1;I^1}.
\end{eqnarray}
Hereafter, we refer to the parameters $\lambda_{I^iJ^i}$ as scale factors.

2. {\it Turning some matrix elements to zero.} For some fixed multi-indexes $N^{3}=N^{3}_0$ and $M^{3}=M^{3}_0$ in (\ref{rho_restore}) we can
find the $\varphi$-parameters such that, along with system (\ref{restore})-(\ref{restore_norm2}), the additional constraint
\begin{eqnarray}
\lambda_{N^{3}_0 M^{3}_0} =0
\end{eqnarray}
is satisfied. This constraint makes the appropriate matrix element vanish: 
\begin{eqnarray}
\rho^{(R;n)}_{N^{3}_0 M^{3}_0} =0.
\end{eqnarray}

3. {\it Rearranging   matrix elements.} For the fixed pair of multi-indexes $N^3=N^3_0$, $M^3=M^3_0$ and $I^3=I^3_0$, $J^3=J^3_0$ in (\ref{gen})  we change formulas (\ref{restore}) as follows:
\begin{eqnarray}\label{restore4}
T_{N^{3}_0M^{3}_0;I^{3}_0J^{3}_0}= \delta_{N^{3}_0J^{3}_0}\delta_{M^{3}_0I^{3}_0} \lambda_{I^{3}_0J^{3}_0} ,\;\;
T_{I^{3}_0J^{3}_0;N^{3}_0M^{3}_0}= \delta_{N^{3}_0J^{3}_0}\delta_{M^{3}_0I^{3}_0} \lambda_{N^{3}_0M^{3}_0} ,\;\;
\end{eqnarray}
and solve system (\ref{restore_norm}), (\ref{restore_norm2}), (\ref{restore4}) for the $\varphi$-parameters.
Then two equations in  (\ref{rho_restore}) which correspond  to this fixed pair must be replaced with
\begin{eqnarray}
\rho^{(R;n)}_{N^{3}_0;M^{3}_0}=\lambda_{I^{3}_0J^{3}_0} \rho^{(S;n)}_{I^{3}_0;J^{3}_0},\;\;
\rho^{(R;n)}_{I^{3}_0;J^{3}_0}=\lambda_{N^{3}_0M^{3}_0} \rho^{(S;n)}_{N^{3}_0;M^{3}_0}.
\end{eqnarray}
Thus, we exchange the elements $\rho^{(S;n)}_{I^{3}_0;J^{3}_0}$ and $\rho^{(S;n)}_{N^{3}_0;M^{3}_0}$.

4. {\it Linear combination of matrix elements with required coefficients.} Since the  elements of $n$-order receiver's coherence matrix $\rho^{(R;n)}$ are linear combinations of the elements of  $\rho^{(S;n)}$, this operation means providing  the required  values to the corresponding elements  of the $T$-operator (\ref{gen})  using  the proper choice of the $\varphi$-parameters. In particular, these combinations can yield the polynomial expansion of certain degree for some functions, see Sec.\ref{Section:lincomb}.

5. {\it Solving  systems of linear algebraic equations}
\begin{eqnarray}\label{Axb}
A \vec x = \vec b,
\end{eqnarray}
where $A$ is a nondegenerate square matrix, $\vec x$ is a vector of unknowns and $\vec b$ is a constant vector. 
We consider the extended receiver (consisting of the two-qubit ancilla and receiver, see Fig.\ref{Fig:CL}) as a single block for applying the unitary transformation establishing the desired operation.  Our protocol is based on  mixed states and  operates as follows.
If there are $M$ elements $\rho^{(S;n)}_{N^{3}_i;M^{3}_i}$, $i=1,\dots, M$ in an $n$-order coherence matrix which map into $\tilde M\le M$ linearly independent combinations at the receiver side (elements of $\rho^{(R;n)}$),  then we can solve a system of $\tilde M$ equations for $\tilde M$ unknowns as follows. Let $\tilde M=M$ for simplicity. 
We first identify   elements $\rho^{(S;n)}_{N^{3}_i;M^{3}_i}$ with the appropriate entries of $\vec b$: 
\begin{eqnarray}
\rho^{(S;n)}_{N^{3}_i;M^{3}_i}=b_i,\;\;i=1,\dots,  M.
\end{eqnarray}
From the other hand, we can write
\begin{eqnarray}
\rho^{(S;n)}_{N^{3}_i;M^{3}_i} = \sum_{j=1}^M A_{ij} x_j,\;\;i=1,\dots, M.
\end{eqnarray}
Accordingly, we can represent  the elements of the receiver's $n$-order coherence matrix as
\begin{eqnarray}\label{rx}
\rho^{(R;n)}_{N^{3}_i;M^{3}_i} = \sum_{j=1}^M \alpha_{ij} x_j,\;\;i=1,\dots,  M,
\end{eqnarray}
where $\alpha_{ij}$, in turn, are some linear combinations of $A_{ij}$ with coefficients depending on $\varphi$-parameters and $t$.
Now we impose the constraints on the coefficients $\alpha_{ij}$:
\begin{eqnarray}\label{alp}
\alpha_{ij} = \alpha \delta_{ij},\;\;\alpha=const,
\end{eqnarray}
which can be satisfied by the proper choice of the $\varphi$-parameters for certain  value of the constant  $\alpha$ at some time instant.
Then eq.(\ref{rx}) yields 
\begin{eqnarray}\label{rx2}
\rho^{(R;n)}_{N^{3}_i;M^{3}_i} = \alpha x_i,\;\;i=1,\dots, M,
\end{eqnarray}
i.e., the scaled values of $x_i$ appear in the entries of the receiver's $n$-order coherence matrix. 
We note that solving system (\ref{alp}) is equivalent to inverting the matrix $A$ with the classical tool. However, been calculated, the $\varphi$-parameters can be used for solving a family of equations with a given matrix $A$ and different vectors $\vec b$.

Similarly, we can consider the  case $\tilde M<M$, see Sec.\ref{Section:solve}.
We notice, that the maximal absolute value of the arbitrary  parameter $\alpha$ is restricted because of  the dispersion during  evolution.

\section{Model}
\label{Section:model}

We consider the communication line {of $N=42$ nodes used} in Ref.\cite{Z_2018}. It consists of the   two-qubit sender $S$ (the first and second nodes), the two-qubit receiver $R$ (the 41st and 42nd nodes) and the two-qubit ancilla (the 39th and 40th nodes). The ancilla and receiver  form the four-node extended receiver $ER$, which is connected to the sender  through the transmission  line $TL$. 
The spin dynamics is governed by the XX-Hamiltonian with the dipole-dipole interaction
\begin{eqnarray}\label{XY}
&&H=\sum_{j>i} D_{ij} (I_{ix}I_{jx} +I_{iy}I_{jy}),\\\label{comm0}
&&[H,I_z]=0,
\end{eqnarray}
where $D_{ij}=\frac{\gamma^2 \hbar}{r_{ij}^3}$ is the coupling constant between 
the $i$th and $j$th nodes, $\gamma$ is the gyromagnetic ratio, $\hbar$ is the Planck constant,  $r_{ij}$ is the distance between the $i$th and $j$th nodes,
$I_{i\alpha}$ ($\alpha=x,y,z$)  is the projection operator of the $i$th spin on the $\alpha$ axis and $I_z=\sum_i I_{iz}$. We also consider the tensor-product initial state (\ref{tpIS}) with an arbitrary $\rho^{(S)}(0)$, while the  subsystem $TL\cup ER$ is in the state without excitations, i.e.,
\begin{eqnarray}
 \label{inTLB2}
\rho^{(TL,ER)}(0) &=&{\mbox{diag}}(1,0,\dots).
 \end{eqnarray}
 Such initial state restricts the evolution of the  spin dynamics  to the  two-excitation subspace which significantly simplifies calculations.
In virtue  of  initial condition (\ref{inTLB2}),  eq.(\ref{T}) with setting 
\begin{eqnarray}
&&
N^3=\{n_1n_2\},\;\;M^3=\{m_1m_2\},\;\;I^1=\{i_1i_2\},\;\;J^1=\{j_1j_2\}, \\\nonumber
&&
\{N^1N^2\}=J_{N-2}=\{j_1\dots j_{N-2}\}, \;\;
\rho^{(TL,ER)}_{I^2I^3;I^2I^3} = \delta_{I^2 0^2} \delta_{I^3 0^3},
\end{eqnarray}
 gets the following form:
\begin{eqnarray}\label{T2}
T_{n_1n_2m_1m_2;i_1i_2j_1j_2} = \sum_{J_{N-2}}  W_{J_{N-2}n_1n_2;i_1i_2 0_{N-2}}  W^+_{j_1j_2 0_{N-2};J_{N-2}m_1m_2},
\end{eqnarray}
where   $0_{N-2}$ is the set of $N-2$ zeros. In this case $W_{J_{N-2}n_1n_2;i_1i_2 0_{N-2}}$ can be considered as  the elements of the Kraus operators \cite{Kraus}, because they satisfy the constraint
\begin{eqnarray}
\sum_{m_1,m_2} \sum_{J_{N-2}} W^+_{j_1j_2 0_{N-2};J_{N-2}m_1m_2}W_{J_{N-2}m_1m_2;i_1i_2 0_{N-2}} = \delta_{i_1j_1}\delta_{i_2j_2},
\end{eqnarray}
which follows from the definition (\ref{W}) of the operator $W$. 

 In our protocol, we use  the expansion of the density matrices in the sums of the multiple-quntum (MQ) coherence matrices \cite{FL} which read in the two-qubit case as
\begin{eqnarray}\label{MQ}
\rho^{(S)} = \sum_{k=-2}^2 \rho^{(S;k)},\;\;\;\rho^{(R)} = \sum_{k=-2}^2 \rho^{(R;k)},
\end{eqnarray}
where  $\rho^{(S;k)}$ and  $\rho^{(R;k)}$ are  the $k$-order  coherence matrices of, respectively, the sender and receiver.  

Following Ref.\cite{Z_2018}, we  introduce notation
\begin{eqnarray}
D_{i(i+1)}\equiv \delta_i,\;\;i=1,\dots,N-1,
\end{eqnarray}
and consider  the  chain 
with the following coupling constants: 
\begin{eqnarray}
 \delta_k=\delta,\;\; 3\le k\le N-3,\;\;\;\delta_1 =\delta_{N-1},  \;\; \delta_2 =\delta_{N-2}.
\end{eqnarray}
Here  $\delta_1$, $\delta_2$ and $t$ are chosen to maximize the intensity of the second-order coherence. For $N=42$, their optimal values read \cite{Z_2018}:
\begin{eqnarray}
\delta_1 =0.3005 \delta, \;\; \delta_2 =0.5311\delta, \;\; 
\delta t=58.9826.
\end{eqnarray}


\section{Operations with nondiagonal  elements of density matrix}
\label{Section:numerics}

In this section, we simulate the operations with density matrix elements discussed in 
Sec.\ref{Section:manipulations}. According to eqs. (\ref{gen}) and (\ref{T2}),
the general structure of the upper nondiagonal elements of the receiver's density matrix  is following: 
\begin{eqnarray}
\label{g2ord}
&&
\rho^{(R;2)}_{00;11}= d \rho^{(S;2)}_{00;11},\\
\label{g1ord1} 
&&
\rho^{(R;1)}_{00;n_1n_2} = a_{n_1n_2;01} \rho^{(S;1)}_{00;01} + a_{n_1n_2;10} \rho^{(S;1)}_{00;10} + 
b_{n_1n_2;01} \rho^{(S;1)}_{01;11} + b_{n_1n_2;10} \rho^{(S;1)}_{10;11},\;\;n_1+n_2=1,\\\label{g1ord2} 
&&
\rho^{(R;1)}_{n_1n_2;11} = c_{n_1n_2;01} \rho^{(S;1)}_{01;11} + c_{n_1n_2;10} \rho^{(S;1)}_{10;11},\;\;n_1+n_2=1,\\\label{g0ord}
&&
\rho^{(R;0)}_{01;10} = \sum_{n_1+n_2=1} f_{n_1n_2;n_1n_2} \rho^{(S;0)}_{n_1n_2;n_1n_2} + f_{01;10} \rho^{(S;0)}_{01;10}+
f_{10;01} \rho^{(S;0)}_{10;01}+f_{11;11} \rho^{(S;0)}_{11;11},
\end{eqnarray}
where
\begin{eqnarray}\label{ddd}
&&
d=W^+_{110_{N-2};0_{N-2}11},\\\nonumber
&&
a_{n_1n_2;i_1i_2} = W^+_{i_1i_20_{N-2};0_{N-2}n_1n_2},\;\;  b_{n_1n_2;i_1i_2} = \sum_{|J_{N-2}|=1} W_{J_{N-2}00;i_1i_2 0_{N-2}} W^+_{110_{N-2};J_{N-2}n_1n_2},\\\nonumber
&&
c_{n_1n_2;i_1i_2} = W_{0_{N-2}n_1n_2;i_1i_2 0_{N-2}} d=a^+_{i_1i_2;n_1n_2} d,\\\nonumber
&&
f_{n_1n_2;m_1m_2}=W_{0_{N-2}01;n_1n_20_{N-2}}W^+_{m_1m_20_{N-2};0_{N-2}10}=a^+_{n_1n_2;01} a_{10;m_1m_2},\;\;n_1+n_2=m_1+m_2=1,\\\nonumber
&&f_{11;11}=\sum_{|J_{N-2}|=1} W_{J_{N-2} 01;110_{N-2}} W^+_{110_{N-2};J_{N-2}10}.
\end{eqnarray}

\subsection{Partial structural restoring of initial sender's state}
In view of the structure of matrix elements (\ref{g2ord})-(\ref{g0ord}),
system (\ref{restore})-(\ref{restore_norm2}) 
for the structural restoring of the upper nondiagonal elements \cite{Z_2018} now reads 
\begin{eqnarray}\label{StrRest}
&& a_{n_1n_2;n_2n_1}= b_{n_1n_2;m_1m_2} =c_{n_1n_2;n_2n_1}= f_{n_1n_2;n_1n_2}=f_{10;01}=f_{11;11}=0,\\\nonumber
&& n_1+n_2=m_1+m_2=1,
\end{eqnarray}
which must be solved for the $\varphi$-parameters.
This system is satisfied if
\begin{eqnarray}\label{PartRestore1}
&&W_{0_{N-2}n_1n_2;n_2n_1 0_{N-2}} =0,\;\;n_1+n_2=1,\\\label{PartRestore2}
&&\sum_{|J_{N-2}|=1} W_{J_{N-2}00 ;m_1m_2 0_{N-2}}  W^+_{11 0_{N-2};J_{N-2}n_1n_2}=0,\;\;m_1+m_2=n_1+n_2=1, \\\label{PartRestore3}
&&
\sum_{|J_{N-2}|=1}
W_{J_{N-2}01 ;11 0_{N-2}} W^+_{11 0_{N-2}; J_{N-1}10}=0,\;\;n_1+n_2=1.
\end{eqnarray}
Then the restored system reads
\begin{eqnarray}
\label{g2ord2}
&&
\rho^{(R;2)}_{00;11}= d \rho^{(S;2)}_{00;11},\\
\label{g1ord12} 
&&
\rho^{(R;1)}_{00;n_1n_2} = a_{n_1n_2;n_1n_2} \rho^{(S;1)}_{00;n_1n_2} ,\;\;n_1+n_2=1,\\\label{g1ord22} 
&&
\rho^{(R;1)}_{n_1n_2;11} = c_{n_1n_2;n_1n_2} \rho^{(S;1)}_{n_1n_2;11},\;\;n_1+n_2=1,\\\label{g0ord2}
&&
\rho^{(R;0)}_{01;10} =  f_{01;10} \rho^{(S;0)}_{01;10}.
\end{eqnarray}
The structural restoring was considered in \cite{Z_2018}, therefore we do not discuss it further.

\subsection{Turning matrix elements to zero}
\label{Section:zero}
It is interesting to note that not any single element of the restored nondiagonal part of $\rho^{(R)}$   can be turned to zero in our model, because 
vanishing of one of the elements in (\ref{g2ord2}) - (\ref{g0ord2})  implies vanishing of  some others. 
We consider three variants of this operation.

1. The coefficients $a_{01;01}$, $c_{01;01}$ and $f_{01;10}$  have the common factor $W^+_{010_{N-2};0_{N-2}01}$ or its conjugate $W_{0_{N-2}01;010_{N-2}}$.
Therefore, if we choose the $\varphi$-parameters
in Eqs.(\ref{g2ord2})-(\ref{g0ord2}) such that 
\begin{eqnarray}\label{zero1}
W_{0_{N-2}01;010_{N-2}}=0,
 \end{eqnarray}
 then 
 \begin{eqnarray}\label{zero11}
 \rho^{(R;1)}_{00;01}=\rho^{(R;1)}_{01;11}=\rho^{(R;1)}_{01;10}=0.
 \end{eqnarray}
{The explicit general form of the described  transformation reads:
\begin{eqnarray}\label{trZero1}
&&\left(\begin{array}{cccc}
\rho^{(S;0)}_{00;00}&\rho^{(S;1)}_{00;01}&\rho^{(S;1)}_{00;10}&\rho^{(S;2)}_{00;11}\cr
(\rho^{(S;1)}_{00;01})^*&\rho^{(S;0)}_{01;01}&\rho^{(S;0)}_{01;10}&\rho^{(S;1)}_{01;11}\cr
(\rho^{(S;1)}_{00;10})^*&(\rho^{(S;0)}_{01;10})^*&\rho^{(S;0)}_{10;10}&\rho^{(S;1)}_{10;11}\cr
(\rho^{(S;2)}_{00;11})^*&(\rho^{(S;1)}_{01;11})^*&(\rho^{(S;1)}_{10;11})^*&\rho^{(S;0)}_{11;11}
\end{array}\right) \to \\\nonumber
&&
\left(\begin{array}{cccc}
\rho^{(R;0)}_{00;00}&0&a_{10;10}\rho^{(S;1)}_{00;10}&d\rho^{(S;2)}_{00;11}\cr
0&\rho^{(R;0)}_{01;01}&0&0\cr
a_{10;10}^*(\rho^{(S;1)}_{00;10})^*&0&\rho^{(R;0)}_{10;10}&c_{10;10}\rho^{(S;1)}_{10;11}\cr
d^*(\rho^{(S;2)}_{00;11})^*&0&c_{10;10}^*(\rho^{(S;1)}_{10;11})^*&\rho^{(R;0)}_{11;11}
\end{array}\right) .
\end{eqnarray}
}

2. Similarly, the coefficients  $a_{10;10}$, $c_{10;10}$ and $f_{01;10}$  have the common  factor $W^+_{100_{N-2};0_{N-2}10}$ or its conjugate $W_{0_{N-2}10;100_{N-2}}$.
 Therefore,
if the $\varphi$-parameters in (\ref{g2ord2})-(\ref{g0ord2})  are such that 
\begin{eqnarray}\label{zero2}
W_{0_{N-2}10;100_{N-2}}=0,
 \end{eqnarray}
 then 
 \begin{eqnarray}\label{zero22}
 \rho^{(R;1)}_{00;10}=\rho^{(R;1)}_{10;11}=\rho^{(R;1)}_{01;10}=0.
 \end{eqnarray}
{The explicit general form of the described  transformation reads:
\begin{eqnarray}\label{trZero2}
&&\left(\begin{array}{cccc}
\rho^{(S;0)}_{00;00}&\rho^{(S;1)}_{00;01}&\rho^{(S;1)}_{00;10}&\rho^{(S;2)}_{00;11}\cr
(\rho^{(S;1)}_{00;01})^*&\rho^{(S;0)}_{01;01}&\rho^{(S;0)}_{01;10}&\rho^{(S;1)}_{01;11}\cr
(\rho^{(S;1)}_{00;10})^*&(\rho^{(S;0)}_{01;10})^*&\rho^{(S;0)}_{10;10}&\rho^{(S;1)}_{10;11}\cr
(\rho^{(S;2)}_{00;11})^*&(\rho^{(S;1)}_{01;11})^*&(\rho^{(S;1)}_{10;11})^*&\rho^{(S;0)}_{11;11}
\end{array}\right) \to \\\nonumber
&&
\left(\begin{array}{cccc}
\rho^{(R;0)}_{00;00}&a_{01;01}\rho^{(S;1)}_{00;01}&0&d\rho^{(S;2)}_{00;11}\cr
a_{01;01}^*(\rho^{(S;1)}_{00;01})^*&\rho^{(R;0)}_{01;01}&0& c_{01;01}\rho^{(S;1)}_{01;11}    \cr
0&0&\rho^{(R;0)}_{10;10}&0\cr
d^*(\rho^{(S;2)}_{00;11})^*&c_{01;01}^*(\rho^{(S;1)}_{01;11})^* &0&\rho^{(R;0)}_{11;11}
\end{array}\right) .
\end{eqnarray}
}
 3. The coefficients 
 $c_{01;01}$ and $c_{10;10}$ are proportional to $d=W^+_{110_{N-2};0_{N-2}11}$. Therefore, if the
 $\varphi$-parameters in  (\ref{g2ord2})-(\ref{g0ord2}) are such that 
 \begin{eqnarray}\label{zero3}
 W^+_{110_{N-2};0_{N-2}11}=0,
 \end{eqnarray}
 then 
 \begin{eqnarray}\label{zero33}
 \rho^{(R;1)}_{01;11}=\rho^{(R;1)}_{10;11}=\rho^{(R;1)}_{00;11}=0.
 \end{eqnarray}
 {The explicit general form of the described  transformation reads:
\begin{eqnarray}\label{trZero3}
&&\left(\begin{array}{cccc}
\rho^{(S;0)}_{00;00}&\rho^{(S;1)}_{00;01}&\rho^{(S;1)}_{00;10}&\rho^{(S;2)}_{00;11}\cr
(\rho^{(S;1)}_{00;01})^*&\rho^{(S;0)}_{01;01}&\rho^{(S;0)}_{01;10}&\rho^{(S;1)}_{01;11}\cr
(\rho^{(S;1)}_{00;10})^*&(\rho^{(S;0)}_{01;10})^*&\rho^{(S;0)}_{10;10}&\rho^{(S;1)}_{10;11}\cr
(\rho^{(S;2)}_{00;11})^*&(\rho^{(S;1)}_{01;11})^*&(\rho^{(S;1)}_{10;11})^*&\rho^{(S;0)}_{11;11}
\end{array}\right) \to \\\nonumber
&&
\left(\begin{array}{cccc}
\rho^{(R;0)}_{00;00}&a_{01;01}\rho^{(S;1)}_{00;01}&a_{10;10}\rho^{(S;1)}_{00;10}&0\cr
a_{01;01}^*(\rho^{(S;1)}_{00;01})^*&\rho^{(R;0)}_{01;01}&f_{01;10}\rho^{(S;0)}_{01;10}& 0  \cr
a_{10;10}^*(\rho^{(S;1)}_{00;10})^*&f_{01;10}^*(\rho^{(S;0)}_{01;10})^*&\rho^{(R;0)}_{10;10}&0\cr
0&0 &0&\rho^{(R;0)}_{11;11}
\end{array}\right) .
\end{eqnarray}
}

 Thus, in the numerical simulations of  {transformations (\ref{trZero1}), (\ref{trZero2}) and (\ref{trZero3})}, we have to solve eqs. 
 (\ref{PartRestore1}) - (\ref{PartRestore3}) and one of  equations, respectively,  (\ref{zero1}), (\ref{zero2}) or 
 (\ref{zero3}) for the $\varphi$-parameters. The solution of this system is not unique.
 Therefore, to reveal the optimal one, we find   
1000 different solutions  
and choose one corresponding to the maximal sum of the absolute values of the non-zero scale factors $\lambda_{I^iJ^i}$ in the nondiagonal elements of the receiver's density matrix.
{Results of numerical calculation of the factors $a_{ij;nm}$, $c_{ij;nm}$, $f_{01;10}$ and $d$  in  transformations (\ref{trZero1}), (\ref{trZero2}) and (\ref{trZero3}) are collected in Table \ref{Table:1}.}
The list of the $\varphi$-parameters of the unitary transformation corresponding to the transformation (\ref{trZero1})  is given in Appendix, the first line in  Table \ref{Table:2}. 
  \begin{table} {
\begin{tabular}{|c|cccccc|}
 \hline
 &$f_{01;10}$
 & $a_{01;01}$&$ a_{10;10}$&$c_{01;01}$&$ c_{10;10}$&$ d$\cr 
 \hline
Eq.(\ref{trZero1}) &   $0$& $0$&$0.9182 e^{- 1.5135 i} $&$0$&$ 0.4393 e^{1.1232i} $&$0.4784e^{-0.3903 i}  $\cr
 Eq.(\ref{trZero2}) &  $0$& $0.6993e^{-1.4322i} $&$0$&$ 0.4129e^{0.0917i}$&$0$&$0.5904 e^{-1.3405i} $\cr
Eq.(\ref{trZero3}) &   $0.3464e^{2.9264i} $& $0.3846e^{-2.9224i} $&$0.9009 e^{0.0040i} $&$0$&$0$&$0$\cr
 \hline
 \end{tabular}
 \caption{\label{Table:1} Numerical coefficients in  Eqs.(\ref{trZero1}), (\ref{trZero2}) and (\ref{trZero3}).
 }}
 \end{table}

\subsection{Rearrangement  of matrix elements}
\label{Section:exchange}
We consider the
rearrangement   of the elements 
$\rho^{(S;1)}_{00;01}$ and $\rho^{(S;1)}_{00;10}$, which implies equations
\begin{eqnarray}\label{reor1}
\rho^{(R;1)}_{00;01} = a_{01;10} \rho^{(S;1)}_{00;10},\;\; \rho^{(R;1)}_{00;10} = a_{10;01} \rho^{(S;1)}_{00;01},
\end{eqnarray}
instead of (\ref{g1ord12}). Therefore we must replace $a_{n_1n_2;n_2n_1}$ with $a_{n_1n_2;n_1n_2}$ in system (\ref{StrRest}), which 
implies
\begin{eqnarray} \label{PartRestore12}
W_{0_{N-2} n_1n_2;n_1n_20_{N-2}}=0
\end{eqnarray}
instead of (\ref{PartRestore1}).
Consequently, according to (\ref{g1ord2}) and (\ref{ddd}), the elements  $\rho^{(S;1)}_{01;11}$ and  $\rho^{(S;1)}_{10;11}$ interchange as well, i.e., we have
\begin{eqnarray}\label{reor2}
\rho^{(R;1)}_{01;11} = c_{01;10} \rho^{(S;1)}_{10;11},\;\; \rho^{(R;1)}_{10;11} = c_{10;01} \rho^{(S;1)}_{01;11}
\end{eqnarray}
instead of (\ref{g1ord22}).
{The explicit general form of the described  transformation reads:
\begin{eqnarray}\label{trRear}
&&\left(\begin{array}{cccc}
\rho^{(S;0)}_{00;00}&\rho^{(S;1)}_{00;01}&\rho^{(S;1)}_{00;10}&\rho^{(S;2)}_{00;11}\cr
(\rho^{(S;1)}_{00;01})^*&\rho^{(S;0)}_{01;01}&\rho^{(S;0)}_{01;10}&\rho^{(S;1)}_{01;11}\cr
(\rho^{(S;1)}_{00;10})^*&(\rho^{(S;0)}_{01;10})^*&\rho^{(S;0)}_{10;10}&\rho^{(S;1)}_{10;11}\cr
(\rho^{(S;2)}_{00;11})^*&(\rho^{(S;1)}_{01;11})^*&(\rho^{(S;1)}_{10;11})^*&\rho^{(S;0)}_{11;11}
\end{array}\right) \to \\\nonumber
&&
\left(\begin{array}{cccc}
\rho^{(R;0)}_{00;00}&a_{01;10}\rho^{(S;1)}_{00;10}&a_{10;01}\rho^{(S;1)}_{00;01}&d\rho^{(S;2)}_{00;11}\cr
a_{01;10}^*(\rho^{(S;1)}_{00;10})^*&\rho^{(R;0)}_{01;01}&f_{01;10}\rho^{(S;0)}_{01;10}& c_{01;10}\rho^{(S;1)}_{10;11}   \cr
a_{10;01}^*(\rho^{(S;1)}_{00;01})^*&f_{01;10}^*(\rho^{(S;0)}_{01;10})^*&\rho^{(R;0)}_{10;10}&c_{10;01}\rho^{(S;1)}_{01;11} \cr
d^*(\rho^{(S;2)}_{00;11})^*&c_{01;10}^*(\rho^{(S;1)}_{10;11})^*  &c_{10;01}^*(\rho^{(S;1)}_{01;11})^*&\rho^{(R;0)}_{11;11}
\end{array}\right). 
\end{eqnarray}
}

The interchange  between the pairs $\rho^{(S;1)}_{00;01}$, $\rho^{(S;1)}_{00;10}$ and $\rho^{(S;1)}_{01;11}$, $\rho^{(S;1)}_{10;11}$ 
is impossible  because of the structure of the appropriate elements  of $\rho^{(R)}$  (see eqs.(\ref{g1ord1}) and (\ref{g1ord2}), there are no elements $\rho^{(S;1)}_{00;01}$ and  $\rho^{(S;1)}_{00;10}$ in the linear combination (\ref{g1ord2})).

Thus, we have to solve system (\ref{PartRestore2}), (\ref{PartRestore3}) and (\ref{PartRestore12}) for the $\varphi$-parameters. Again, the solution is not unique. Therefore, we find  1000 different solutions and
 choose one corresponding to the maximal sum of the absolute values of  the scale factors $\lambda_{I^iJ^i}$ in the  nondiagonal elements of the density matrix.
The result of numerical { calculations yields    the  following values for the factors  $a_{ij;nm}$, $c_{ij;nm}$, $f_{01;10}$ and $d$ in transformation (\ref{trRear}):
\begin{eqnarray}\label{acfd}
\begin{array}{lll}
f_{01;01}=0.2028 e^{1.2890 i}, & a_{01;10}=0.9175 e^{0.0427i},& a_{10;01}= 0.2210 e^{1.3317i}\cr
c_{01;10}=0.4095 e^{1.6726i}, & c_{10;01}=  0.0987e^{ 0.3835i}, & d=0.4464 e^{1.7153i}.
\end{array}
\end{eqnarray}
}
The list of the $\varphi$-parameters of the unitary transformation corresponding to { rearrangement (\ref{trRear})}  is given in Appendix, the second line in  Table \ref{Table:2}.

\subsection{Linear  combinations of matrix elements}
\label{Section:lincomb}
The structure of  matrix  elements (\ref{g2ord}) - (\ref{g0ord}) shows  that,
 among the elements of the first order coherence matrix, only $\rho^{(R;1)}_{00;01}$ and $\rho^{(R;1)}_{00;10}$ can be two indepdendent linear combinations of all the elements  of $\rho^{(S;1)}$ with arbitrary coefficients. Also the coefficient in $\rho^{(R;2)}_{00;11}$ (second-order coherence matrix) is independent on the coefficients in 
$\rho^{(R;1)}_{00;01}$ and $\rho^{(R;1)}_{00;10}$. But the element $\rho^{(R;0)}_{01;10}$ has a different feature. Only the coefficient $f_{11;11}$ is independent on the other coefficients in $\rho^{(R;1)}_{00;01}$, $\rho^{(R;1)}_{00;10}$ and $\rho^{(R;2)}_{00;11}$.
Thus, we can arrange the following independent linear combinations:
\begin{eqnarray}\label{lin1}
&&
\rho^{(R;1)}_{00;01}=\alpha_{1} \rho^{(S;1)}_{00;01}+\alpha_{2} \rho^{(S;1)}_{00;10}+\alpha_{3} \rho^{(S;1)}_{01;11}+\alpha_{4} \rho^{(S;1)}_{10;11},\\\label{lin2}
&&
\rho^{(R;1)}_{00;10}=\beta_{1} \rho^{(S;1)}_{00;01}+\beta_{2} \rho^{(S;1)}_{00;10}+\beta_{3} \rho^{(S;1)}_{01;11}+\beta_{4} \rho^{(S;1)}_{10;11},\\\label{lin3}
&&
\rho^{(R;2)}_{00;11}=\gamma \rho^{(S;2)}_{00;11},
\end{eqnarray}
where $\alpha_i$, $\beta_i$ and $\gamma$ are arbitrary    scalar coefficients (but such that  $\rho^{(R)}$ remains the nonnegatively definite matrix with unit trace). 

In particular, if 
\begin{eqnarray}\label{xxx}
\rho^{(S;1)}_{00;01}=x,\;\; \rho^{(S;1)}_{00;10}=x^2,\;\;\rho^{(S;1)}_{01;11}=x^3,\;\; \rho^{(S;1)}_{10;11}=x^4,\;\;\rho^{(S;2)}_{00;11}=x^5,
\end{eqnarray}
the above combinations represent the power series in $x$.
For instance, let $\varphi$-parameters satisfy the system 
\begin{eqnarray}\label{comb1}
&&
\alpha_{1}=\alpha,\;\; \alpha_2= \frac{\alpha}{2},\;\; \alpha_3= \frac{\alpha}{3!},\;\; \alpha_4= \frac{\alpha}{4!},\\\nonumber
&&
\beta_{1}=-\alpha,\;\;\beta_2= \frac{\alpha}{2},\;\; \beta_3= - \frac{\alpha}{3!},\;\; \beta_4=  \frac{\alpha}{4!},\;\;\gamma=\frac{\alpha}{5!}, \;\; \alpha=const
\end{eqnarray}
then we have {the 4-degree polynomial}  expansions of $\alpha(e^x-1)$ and $\alpha(e^{-x}-1)$ in the entries  $\rho^{(R;1)}_{00;01}$ and $\rho^{(R;1)}_{00;10}$ respectively, while the entry 
$\rho^{(R;1)}_{00;10}$ yields a fifth-power term in the expansion of $\alpha(e^x-1)$.
{The  general form of the described transformation for real $x$ and $\alpha$ reads:
\begin{eqnarray}\label{trLin}
&&
\left(\begin{array}{cccc}
\rho^{(S;0)}_{00;00}&x&x^2&x^5\cr
x&\rho^{(S;0)}_{01;01}&\rho^{(S;0)}_{01;10}&x^3\cr
x^2&(\rho^{(S;0)}_{01;10})^*&\rho^{(S;0)}_{10;10}&x^4\cr
x^5&x^3&x^4&\rho^{(S;0)}_{11;11}
\end{array}\right) \to 
\\\nonumber
&&
\left(\begin{array}{cccc}
\rho^{(R;0)}_{00;00}(x)&\alpha\left(x+\frac{x^2}{2}+\frac{x^3}{3!}+ \frac{x^4}{4!}\right)&\alpha\left(-x+\frac{x^2}{2}-\frac{x^3}{3!}+ \frac{x^4}{4!}\right)     &\alpha\frac{x^5}{5!}\cr
\alpha\left(x+\frac{x^2}{2}+\frac{x^3}{3!}+ \frac{x^4}{4!}\right)
&\rho^{(R;0)}_{01;01}&\rho^{(R;0)}_{01;10}&\rho^{(R;1)}_{01;11} (x)  \cr
\alpha\left(-x+\frac{x^2}{2}-\frac{x^3}{3!}+ \frac{x^4}{4!}\right) &(\rho^{(R;0)}_{01;10})^*&\rho^{(R;0)}_{10;10}&\rho^{(R;1)}_{10;11}(x) \cr
\alpha\frac{x^5}{5!}&(\rho^{(R;1)}_{01;11}(x))^*  &(\rho^{(R;1)}_{10;11}(x))^*&\rho^{(R;0)}_{11;11}
\end{array}\right),
\end{eqnarray}
where $\rho^{(R;1)}_{ij;nm}(x)$ are some polynomial in $x$ which we  do not write explicitly.
}
In the numerical simulations,  we find  1000 different solutions of system (\ref{comb1}) and choose one that  maximizes $|\alpha|$, which yields $|\alpha|_{max}=0.5399$.
The list of $\varphi$-parameters of the unitary transformation corresponding to this manipulation   is given in Appendix, the third line in  Table \ref{Table:2}.

\subsection{Solving system of linear algebraic equations}
\label{Section:solve}

We show that the 2$\times$ 2 linear algebraic system (\ref{Axb}) with real coefficients
\begin{eqnarray}\label{linsyst}
\left(\begin{array}{cc}
a_{11}&a_{12}\cr
a_{21}&a_{22}
\end{array}\right)   \left(\begin{array}{c}
x_1\cr
x_2
\end{array}\right) =   \left(\begin{array}{c}
b_1\cr
b_2
\end{array}\right)   
\end{eqnarray}
can be solved using the tool of unitary transformations on the extended receiver.
Let 
\begin{eqnarray}
&&
\rho^{(S;1)}_{00,01}=b_1,\;\;\rho^{(S;1)}_{00,10}=b_2,\\\nonumber
&&
\rho^{(S;1)}_{01,11}=\rho^{(S;1)}_{10,11}=0.
\end{eqnarray}
We also assume that 
\begin{eqnarray}
\rho^{(S;1)}_{00,01} = a_{11} x_1 + a_{12} x_2,\;\;\rho^{(S;1)}_{00,10} = a_{21} x_1 + a_{22} x_2.
\end{eqnarray}
After evolution {under} the operator $W$ we have
\begin{eqnarray}\label{rho_alg}
&&\rho^{(R;1)}_{00;01} = \alpha_{11} x_1 + \alpha_{12} x_2,\\\nonumber
&&\rho^{(R;1)}_{00;10} = \alpha_{21} x_1 + \alpha_{22} x_2,
\end{eqnarray}
where $\alpha_{ij}$ depend on the $\varphi$-parameters and on the parameters $a_{ij}$ of linear system (\ref{linsyst}).
We choose the $\varphi$-parameters to satisfy the conditions
\begin{eqnarray}\label{sollin}
\alpha_{12}=\alpha_{21}=0,\;\;
\alpha_{11}=\alpha_{22}=c=const.
\end{eqnarray}
Then system (\ref{rho_alg}) gets the following form
\begin{eqnarray}\label{rho_alg2}
&&\rho^{(R;1)}_{00;01} = c x_1 ,\\\nonumber
&&\rho^{(R;1)}_{00;10} = c x_2.
\end{eqnarray}
Thus, the elements $\rho^{(R;1)}_{00;01}$ and $\rho^{(R;1)}_{00;10}$   of the first-order coherence matrix represent solution of  original system (\ref{linsyst}) up to the factor $c$.
{The general  form of the described  transformation reads:
\begin{eqnarray}\label{trSol}
\left(\begin{array}{cccc}
\rho^{(S;0)}_{00;00}&b_1&b_2&\rho^{(S;2)}_{00;11}\cr
b_1&\rho^{(S;0)}_{01;01}&\rho^{(S;0)}_{01;10}&0\cr
b_2&(\rho^{(S;0)}_{01;10})^*&\rho^{(S;0)}_{10;10}&0\cr
(\rho^{(S;2)}_{00;11})^*&0&0&\rho^{(S;0)}_{11;11}
\end{array}\right) \to 
\left(\begin{array}{cccc}
\rho^{(R;0)}_{00;00}&x_1&x_2&\rho^{(R;2)}_{00;11}\cr
x_1&\rho^{(R;0)}_{01;01}&\rho^{(R;0)}_{01;10}&\rho^{(R;1)}_{01;11}(x)\cr
x_2&(\rho^{(R;0)}_{01;10})^*&\rho^{(R;0)}_{10;10}&\rho^{(R;1)}_{10;11}(x)\cr
(\rho^{(R;2)}_{00;11})^*&(\rho^{(R;1)}_{01;11}(x))^*&(\rho^{(ë;1)}_{10;11}(x))^*&\rho^{(R;0)}_{11;11}
\end{array}\right) ,
\end{eqnarray}
where $\rho^{(R;1)}_{ij;nm}(x)$ are some linear functions of $x$ which we  do not write explicitly.}
As an example, we take $a_{11}=0.4$, 
$a_{12}=0.3$, $a_{21}=0.6$, $a_{22}=0.2$ and solve  system (\ref{sollin}) for the $\varphi$-parameters. Again, the  solution is not unique. We find the solution corresponding to the maximal possible value of the real positive $c$ in  (\ref{rho_alg2}), $c_{max}=0.1094$. 
The list of $\varphi$-parameters of the unitary transformation corresponding to this operation is given in Appendix, the 4th line in  Table \ref{Table:2}.

\section{Conclusions}
\label{Section:conclusions}

The unitary transformations on the extended receiver can be an effective tool not only for structural restoring of the transferred state (as  shown in Ref.\cite{Z_2018}) but also for realizing a set of other manipulations with  elements of the receiver's density matrix, { such as turning some matrix elements to zero (Sec.\ref{Section:zero}), exchanging the positions of some elements (Sec.\ref{Section:exchange}), constructing certain linear combinations of the elements (Sec.\ref{Section:lincomb}) and solving the systems of linear algebraic equations (Sec.\ref{Section:solve}).}   
{ Being larger than the receiver, the extended receiver admits larger number of  free parameters ($\varphi$-parameters in this paper)  in the applied unitary transformation, and these  parameters can be effectively involved in  the above structural manipulations with the mixed receiver's state.}  This is another benefit of the extended receiver which first was used to  improve the characteristics of the communication line in the protocol of remote state creation \cite{BZ_2016}. It is important that our protocols  do not use a classical channel. 

We shall notice that certain  restrictions in manipulation with matrix elements in Sec.\ref{Section:numerics} in comparison with general formulas proposed in Sec.\ref{Section:manipulations} (for instance, only triplets of matrix elements can be turned to zero rather than any particular element, 
restricted possibilities for rearranging the matrix elements)
are associated with the { particularly chosen  initial state of the transmission line and receiver (\ref{inTLB2}), which is the state without  excitations.} Using a different initial state (for instance, the thermal equilibrium one) we might extend possibilities of such manipulations.

The considered  operations with matrix elements via the unitary transformations on  the extended receiver  represent  simplest examples of quantum realization of algebraic operations. 
The further development of this approach with the purpose to simulate quantum counterparts for
more complicated algebraic algorithms as well as methods of realizations of appropriate unitary transformations is of interest.

This work is performed in accordance with the state task of the
Ministry of Education and Science of the Russian Federation, state
registration No. 0089-2019-0002, and by the Russian Foundation for Basic Research (Grants No.20-03-00147).

 \section{Appendix: Unitary transformation as a tool for  manipulations with matrix elements}
 \label{Section:appendix}
 For the matrix representation of the unitary transformation $U^{(ER)}$ and density matrix $\rho^{(ER)}$ we use the basis of eigenstates of $I^{(ER)}_z$,
 \begin{eqnarray}\nonumber
 |0000\rangle, \;\;|0001\rangle, \;\;|0010\rangle,\;\;|0011\rangle, \;\;|0100\rangle, \;\;|0101\rangle,\;\; 
 |0110\rangle, \;\;|1000\rangle, \;\;|1001\rangle, \;\;|1010\rangle, \;\;|1100\rangle.
\end{eqnarray}
The unitary operator $U^{(ER)}$ can be written in the form
 \begin{eqnarray}\label{V0}
 U^{(ER)}=\prod_{n=1}^{11}\prod_{m>n} e^{i \varphi^{(2)}_{n,m} \gamma^{(2;nm)}} e^{i \varphi^{(1)}_{n,m}\gamma^{(1;nm)}},
 \end{eqnarray}
 where  $\varphi^{(k)}_{n,m}$ are scalar parameters, the product is  ordered in such a way that $n$ and $m$ 
 increase from the right to the left and the Hermitian matrices $\gamma^{(1;ij)}$, $\gamma^{(2;ij)}$, $j>i$ 
 have the following nonzero elements \cite{Z_2018}:
 \begin{eqnarray}
 \label{bases}
 &&
 \gamma^{(1;ij)}_{ij}= \gamma^{(1;ij)}_{ji}=1,\;\;\;\gamma^{(2;ij)}_{ij}= -\gamma^{(1;ji)}_{ji}=-i,\\\label{nonzerovarphi}
 (i,j)&\in&\{(2,3),(2,5),(2,8),(3,5), (3,8), (4,6), (4,7), (4,9), (4,10), (4,11), (5,8), \\\nonumber
 &&(6,7), (6,9), (6,10), (6,11), (7,9), (7,10), (7,11), (9,10), (9,11), (10,11)\}.
 \end{eqnarray}

 In Table \ref{Table:2}, we represent the families  of $\varphi$-parameters for the unitary transformations corresponding to the manipulations with matrix elements discussed in Secs.\ref{Section:zero}-\ref{Section:solve}. The first, second and third  lines in this table correspond, respectively,  to  transformation (\ref{zero11}),  to rearrangements (\ref{reor1}),(\ref{reor2}) and  to linear combination (\ref{lin1}) - (\ref{comb1}). The parameters of the unitary transformation solving the system of two linear  algebraic equations (\ref{linsyst}) are collected in the fourth line of this table.
 \begin{table}[h]
\begin{tabular}{|c|ccccccccccccc|}
 \hline
 $\#$&$\varphi^{(1)}_{2,3}$&$\varphi^{(1)}_{2,5}$&$\varphi^{(1)}_{2,8}$&$\varphi^{(1)}_{3,5}$&$\varphi^{(1)}_{3,8}$&$\varphi^{(1)}_{4,6}$& $\varphi^{(1)}_{4,7}$& $\varphi^{(1)}_{4,9}$& $\varphi^{(1)}_{4,10}$& $\varphi^{(1)}_{4,11}$& $\varphi^{(1)}_{5,8}$& $\varphi^{(1)}_{6,7}$& $\varphi^{(1)}_{6,9}$\cr 
 \hline
 1.&0.2871&1.2703&5.3837&0.0157&4.4081&4.3842&1.4712& 0.7807&5.7679&1.4273&5.7371&5.7107&5.0750\cr 
 2.&6.2612&1.3199&4.6848&2.7895&3.1939&4.9799&1.0863&3.7417&0.3186&1.2688&0.1286&4.5929&0.1008\cr 
 3.&1.4280&1.1782&3.5900&0.3906&0.8475&1.1926&5.9163&5.5803&4.6112&5.8579&4.0289&3.5539&1.9443\cr 
 4.&3.7020&3.5217&3.9665&5.0063&2.3526&5.2017&1.0258&0.3117&1.0826&1.5443&3.9441&0.9333&2.0108\cr 
 \hline
 \end{tabular}
 \begin{tabular}{|c|ccccccccccccc|}
 \hline
 $\#$&$\varphi^{(2)}_{2,3}$&$\varphi^{(2)}_{2,5}$&$\varphi^{(2)}_{2,8}$&$\varphi^{(2)}_{3,5}$&$\varphi^{(2)}_{3,8}$&$\varphi^{(2)}_{4,6}$& $\varphi^{(2)}_{4,7}$& $\varphi^{(2)}_{4,9}$& $\varphi^{(2)}_{4,10}$& $\varphi^{(2)}_{4,11}$& $\varphi^{(2)}_{5,8}$& $\varphi^{(2)}_{6,7}$& $\varphi^{(2)}_{6,9}$\cr 
 \hline
 1.&1.5038&6.3390&3.6119&3.4403&1.5498&1.6824&1.7554&4.9749&3.0629&1.6850&0.8209&3.3091&4.7280\cr 
 2.&3.2009&4.7175&4.7844&4.2364&4.4368&2.3704&4.7478&3.8223&0.0666&4.3057&2.4175&3.7802&5.0880\cr 
 3.&3.5958&4.7751&3.3132&5.7055&5.2464&5.7296&3.2206&6.1773&2.7911&1.8675&1.1603&0.2330&2.7525\cr 
 4.&5.0184&3.4904&3.8684&5.7569&6.2019&4.5391&3.6655&3.9450&0.5715&6.2573&5.9570&3.0617&2.9621\cr 
 \hline
 \end{tabular}
 \begin{tabular}{|c|cccccccc|}
 \hline
 $\#$&$\varphi^{(1)}_{6,10}$&$\varphi^{(1)}_{6,11}$&$\varphi^{(1)}_{7,9}$&$\varphi^{(1)}_{7,10}$&$\varphi^{(1)}_{7,11}$&$\varphi^{(1)}_{9,10}$& $\varphi^{(1)}_{9,11}$& $\varphi^{(1)}_{10,11}$\cr 
 \hline
 1.&4.2237&5.9065&2.8883&2.5863&5.3478& 0.8597&4.3707&1.4844\cr 
 2.&3.7829&3.0648&0.4342&0.1331&1.6031&5.1080&4.3414&4.7175\cr
 3.&2.3509&0.2806& 2.1732&2.5927&0.7668&2.1064&0.4130&5.7157\cr
 4.&3.9155&4.5563&3.5955&3.1836&4.6770&2.1579&4.3653& 1.0752\cr
 \hline
 \end{tabular}
 \begin{tabular}{|c|cccccccc|}
 \hline
 $\#$&$\varphi^{(2)}_{6,10}$&$\varphi^{(2)}_{6,11}$&$\varphi^{(2)}_{7,9}$&$\varphi^{(2)}_{7,10}$&$\varphi^{(2)}_{7,11}$&$\varphi^{(2)}_{9,10}$& $\varphi^{(2)}_{9,11}$& $\varphi^{(2)}_{10,11}$\cr 
 \hline
 1.&4.2920&2.5717&0.7667&0.2932&5.0036&5.2766&6.1909&2.5412\cr 
 2.&5.0943&0.0743&1.8950&0.8286&5.3777&2.7487&1.9213&2.3468\cr
 3.&5.8947&1.2263&0.1978&2.4723&3.5912&1.2593&1.2938&2.9502\cr
 4.&4.5605&2.3953&3.0425&2.0803&4.7866&4.6980&0.8416&4.1266\cr
 \hline
 \end{tabular}
 \caption{\label{Table:2} Families  of the $\varphi$-parameters of unitary transformation (\ref{V0}) corresponding to the manipulations with the receiver's density matrix elements discussed in Secs.\ref{Section:zero} - \ref{Section:solve}. The lines  1 -- 4  correspond, respectively, to 
  the transformation (\ref{zero11}), rearrangements (\ref{reor1}) and (\ref{reor2}), linear combination of elements (\ref{lin1}) - (\ref{comb1}) and solving  the system of two linear  algebraic equations (\ref{linsyst}). 
  }
 \end{table}

\end{document}